# Indoor Massive MIMO: Uplink Pilot Mitigation Using Channel State Information Map

Ahmad Abboud[1], Ali H. Jaber[2]
[2]Department of Statistics
[2]Lebanese University
[2]Nabatieh, Lebanon
[1]Ahmad.Abboud@etu.unilim.fr, [2]ali.jaber@ul.edu.lb

Jean-Pierre Cances [3], Vahid Meghdadi[4]
[1,3,4]Department C2S2
[1,3,4]LimogesUniversity
[1,3,4]Limoges, France
[3,4]{cances, meghdadi}@ensil.unilim.fr

*Abstract*—**Massive MIMO brings both motivations and challenges to develop the 5[th] generation Mobile wireless technology. The promising number of users and the high bitrate offered per unit area are challenged by uplink pilot contamination due to pilot reuse and a limited number of orthogonal pilot sequences.**

**This paper proposes a solution to mitigate uplink pilot contamination in an indoor scenario where multi-cell share the same pool of pilot sequences, that are supposed to be less than the number of users. This can be done by reducing uplink pilots using Channel State Information (CSI) prediction.**

**The proposed method is based on machine learning approach, where a quantized version of Channel State Information (QCSI) is learned during estimation session and stored at the Base Station (BS) to be exploited for future CSI prediction. The learned QCSI are represented by a weighted directed graph, which is responsible to monitor and predict the CSI of User Terminals (UTs) in the local cell.**

**We introduce an online learning algorithm to create and update this graph which we call CSI map. Simulation results show an increase in the downlink sum-rate and a significant feedback reduction.**

## I. INTRODUCTION

The incredible increase in the number of users that need to be connected with their mobile equipment's to the Internet and the aggressive data transfer between UTs, demand the development of new generation of wireless networks with higher capacity. Massive MIMO offers a significant gain in achievable sum-rate using spatial multiplexing [1]. In multiple cell scenario, this achievable rate degrades due to the contamination effect of the uplink pilots from UTs of neighbor cells [2]. Researches on Massive MIMO had shown that increasing the number of antennas at the base station (BS) will reduce the fast fading effect and the noise effect while the pilot contamination effect persists [3], [4].

According to [1], It is recognized that the acquisition of channel knowledge is facilitated by time-division duplex (TDD) operation, where UTs need to transmit their pilots on the reverse link to allow channel estimation at the base station (BS). Under high mobility conditions, there is no enough time before the channel changes to transmit reverse pilots and then transmit the forward data streams. The short coherence interval leads to a limited number of orthogonal pilot sequences to be used within the UTs. Thus, pilot reuse may take place, which leads to uplink pilot contamination.

In an indoor scenario, UTs and scattering objects are less mobile, which leads to a smoother change in channel state information compared to an outdoor scenario. This fact can increase channel coherence time for most UTs.

Quantized channel state information was studied by [5], the authors propose a vector-quantization approach to channel state information encoding, which requires modest feedback bit rate. Another patent was published by the same authors on quantized channel information prediction in multiple antenna systems [6]. In this paper, we use a quantized CSI (QCSI) based on vector quantization of geometric attenuation and shadow fading parameters.

CSI prediction can be modeled based (example [7]) or non-model based (example [8] using stochastic channel model). In [9] adaptive codebook geodesic based channel prediction is proposed, where simulation results show that the proposed scheme can effectively mitigate the feedback delay and clustering even with only a 4-bit codebook. CSI prediction at the base station was recently studied by [10], where the BS selects the set of user terminals that will exhibit pilot training. However, our proposed technique allows the UT to decide whether to send its pilot or not. We follow the technique used in [8] to predict the CSI.

This paper proposes a technique to reduce uplink pilot feedback by predicting CSI at the BS instead of estimating CSI for each uplink session. To perform this work we introduce two types of TDD formats, a predictive format that does not include pilots and an initiative format which is similar to conventional Massive MIMO TDD format. UTs decide which format to use based on the previous received SNR level. On high SNR, a predictive TDD format will be uploaded whilst initiative format will be uploaded at a low level for previously received SNR. At the base station, a quantized version of CSI will be learned and stored in connected nodes as a map that will be used later for channel prediction and precoding. The initiative TDD formats will be exploited as a learning session to update the CSI Map while predictive TDD formats demand the prediction of CSI of a specific UT. Such a technique will have a major impact on mitigating uplink pilot contamination and increasing system sum-rate.

The intuition behind the CSI map was inspired based on (hypothesis1 [11]), which was used for localization and proved by real measurements. We suppose that in each cell there exist a set of UTs that hold their positions for several coherence intervals. This makes sense in the indoor scenario due to limited space. Up to our knowledge, we are the first to introduce CSI map in the indoor Massive MIMO scheme.

The following points can summarize the work achieved by this paper:

- A proposed TTD reciprocity format that enables a reverse link without pilots.
- A learning scheme to create and update the CSI map based on estimated and quantized CSI.
- Finally, simulation results that show the performance of implementing CSI map.

The rest of this paper is organized as follows:

Next, we will present the system model, then a proposed reciprocity scheme will be presented. CSI map will be presented in section IV and then the used CSI quantization technique will be presented in section V. The proposed CSI map learning algorithm will be presented in section VI where numerical results will be discussed in section VII. At the end, we conclude our work in section VIII.

*Notations:* In this paper, $(.)^T$, $(.)^H$ denote transpose and Hermitian transpose, respectively. $(.)^*$ denote the conjugate, det(A) denote the determinant of A, $\odot$ denote element-wise multiplication and $\|A\|$ denote the Frobenius norm.

## II. SYSTEM MODEL

We consider a system of $L$ cells, each cell served by one BS holding $M$ antennas and $K$ UTs each equipped with a single antenna. Assuming a TDD acquisition of channel knowledge, where CSI is estimated or predicted at the BS. During normal channel estimation phase, all users from the $L$ cells uploads their assigned pilot sequences. Considering a worst-case scenario, where synchronized transmission is assumed in all cells [3], the received signal at the j[th] BS will be presented as following:

$$\mathbf{Y}_j^{up} = \sqrt{P_u} \sum_{l=1}^{L} \mathbf{G}_{jl} \mathbf{X}_l^{up} + \mathbf{W}_j^{up} \qquad (1)$$

Where $\mathbf{G}_{jl}$ is the $M \times K$ channel matrix between the j[th] BS containing $M$ antennas and the $K$ UTs in the l[th] cell, i.e. $g_{mk} \triangleq [\mathbf{G}]_{m,k}$ is the channel coefficient between the $m$th antenna of the BS and the $k$th user. $\mathbf{X}_l^{up}$ is the $K \times 1$ symbols vector simultaneously transmitted by the $K$ users in the l[th] cell, $P_u$ is the normalized received SNR of each user by the BS and $\mathbf{W}_j^{up}$ is the $M \times 1$ matrix which represents additive AWGN i.i.d noise vector with zero-mean, unit-variance and $CN$ (0,1). The coefficient $g_{mk}$ can be written as:

$$g_{mk} = h_{mk}\sqrt{\beta_k} \qquad m=1,2,\dots,M \qquad (2)$$

Where $h_{mk}$ is the fast fading coefficient from the $k$th UT to the $m$th antenna. $\sqrt{\beta_k}$ models the geometric attenuation and shadow fading which is assumed to be independent over $m$ and to be constant over many coherent time intervals and known prior. From (2) we obtain:

$$\mathbf{G} = \mathbf{H}\mathbf{D}^{1/2} \qquad (3)$$

Where $\mathbf{H}$ is the $M \times K$ matrix of fast fading coefficients between the $K$ users and the $M$ antennas of the BS, i.e. $h_{mk} \triangleq [\mathbf{H}]_{m,k}$ and $\mathbf{D}$ is the $K \times K$ diagonal matrix, where: $[\mathbf{D}]_{k,k} = \beta_k$ presents the large-scale fading between BS and user k.

Therefore, (1) can be written as:

$$\mathbf{Y}_j^{up} = \sqrt{P_u} \sum_{l=1}^{L} \mathbf{H}_{jl} \mathbf{D}_{jl}^{1/2} \mathbf{X}_j^{up} + \mathbf{W}_j^{up}$$

$$\mathbf{Y}_j^{up} = \underbrace{\sqrt{P_u}\mathbf{H}_{jj}\mathbf{D}_{jj}^{1/2}\mathbf{X}_l}_{prefered\ signal} + \underbrace{\sqrt{P_u}\sum_{l=1,l\neq j}^{L}\mathbf{H}_{jl}\mathbf{D}_{jl}^{1/2}\mathbf{X}_j^{up}}_{contaminated\ signal}$$

$$+ \underbrace{\mathbf{W}_j^{up}}_{noise\ vector} \quad (4)$$

Considering the uplink pilot session of length $\tau \times 1$, then the received signal at the $j^{th}$ BS in the $l^{th}$ cell will be presented as follows:

$$\mathbf{Y}_j^p = \sqrt{\tau P_u}\sum_{l=1}^{L} \mathbf{G}_{jl}[\mathbf{X}_l^p \odot \mathbf{S}_l] + \mathbf{W}_j^{up} \quad (5)$$

Where $\mathbf{S}$ is a $K \times 1$ binary matrix with elements $s_k \in \{0,1\}$, representing by $s_k = 0$; in this case UT that uploads a predictive TDD format (without pilots) and by $s_k = 1$, in this case, UT uploads an initiative TDD format (using pilots). $\mathbf{X}_l^p$ is the $K \times 1$ matrix with elements $x_{lk}^p$, each represents a pilot sequence uploaded from the $k^{th}$ UT of the $l^{th}$ cell.

Let $K' \leq K$, be the set of UTs that uploads their pilots in each cell and $\alpha = \frac{K'}{K}$, represents the probability of a UT in cell $l$ to send an initiative TDD format. Considering the same pilot sequence is reused once in each cell, $L' = r(L \times \alpha)$ will represent the number of cells that upload the same pilot sequence, where $r(.)$ is a function that rounds to the nearest integer.

(6) Gives the Least Squares Estimation of the channel matrix at the jth BS:

$$\widehat{\mathbf{G}}_j = \arg\min_{\mathbf{G}_{jl}} \left\|\frac{1}{\sqrt{P_u}}\mathbf{Y}_j^p - \mathbf{G}_{jl}\mathbf{X}_j^{pH}\right\|^2 \quad (6)$$

The solution of (6) can be expressed as (7):

$$\widehat{\mathbf{G}}_j = \sqrt{\tau P_u}\mathbf{G}_{jj} + \sqrt{\tau P_u}\sum_{l=1,l\neq j}^{L'} \mathbf{G}_{jl} + \widehat{\mathbf{W}}_j^{up} \quad (7)$$

Where $\widehat{\mathbf{W}}_j^{up}$ AWGN still has i.i.d distribution with zero-mean, unit-variance and $CN(0,1)$.

At the Forward link, the $j^{th}$ BS transmits a precoded matrix to all $K$ UTs based on the estimated version of (7). Considering the use of eigen-beamforming linear precoder, the received signal at the $K$ UT antennas can be represented as (8):

$$\mathbf{Y}_j^d = \sqrt{P_d}\sum_{l=1}^{L} \mathbf{G}_{jl}^T \widehat{\mathbf{G}}_j^* \mathbf{X}_j^d + \mathbf{W}_j^d \quad (8)$$

$\mathbf{X}_j^d$ is the $K \times 1$ symbols vector received by the $K$ users in the $l^{th}$ cell, $P_d$ is the normalized received SNR of at each UT and $\mathbf{W}_j^d$ is the $K \times 1$ matrix represents additive AWGN i.i.d noise vector with zero-mean, unit-variance and $CN(0,1)$.

From (7) and (8) we can obtain:

$$\mathbf{Y}_j^d = \sqrt{P_d}\sum_{l=1}^{L} \mathbf{G}_{jl}^T \left[\sqrt{\tau P_u}\sum_{l=1}^{L'} \mathbf{G}_{jl} + \widehat{\mathbf{W}}_j^{up}\right]^* \mathbf{X}_j^d + \mathbf{W}_j^d \quad (9)$$

According to [12] as $M \gg K$ the following relation holds

$$\left(\frac{\mathbf{G}_{jl}^H \mathbf{G}_{jl}}{M}\right)_{M \gg K} = \mathbf{D}_j^{1/2}\left(\frac{\mathbf{H}_{jl}^H \mathbf{H}_{jl}}{M}\right)_{M \gg K} \mathbf{D}_j^{1/2}$$

$$\approx \mathbf{D}_j^{1/2} \quad (10)$$

And $\quad \frac{1}{M}\mathbf{H}_{jl}^T \mathbf{H}_{jl}^* = \mathbf{I}_K \delta_{jl}$

Where $\mathbf{I}_K$ is an Identity matrix with dimension $K \times K$ and $\delta_{jl}$ corresponds to the covariance factor of $\mathbf{H}_{jl}$.

From (9) and (10) we can write the received signal at the $k^{th}$ UT in the $j^{th}$ cell as:

$$\frac{1}{M\sqrt{P_u P_d}} y_{jk}^d = \beta_{jjk} x_{jk}^d + \sum_{l=1,l\neq j}^{L'} \beta_{jlk} x_{jk}^d$$

Where $\beta_{jlk}$ corresponds to the large-scale fading between the $k^{th}$ UT in the $l^{th}$ cell and the $j^{th}$ BS, $x_{jk}$ is the $k^{th}$ element of symbol vector $\mathbf{X}_j^d$. The signal to interference noise ratio of each UT can be written as:

$$SINR = \frac{\beta_{jjk}^2}{\sum_{l=1,l\neq j}^{L'} \beta_{jlk}^2} \quad (11)$$

III. PROPOSED RECIPROCITY SCHEME

We propose two types of TDD protocol format (see figure 1), initiative and predictive formats.

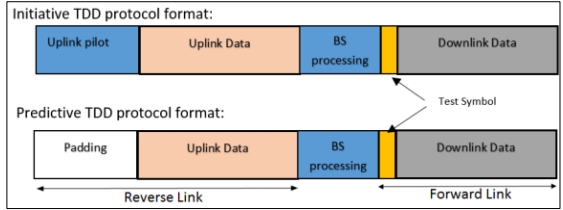

*Figure 1 TDD protocol format*

At each transmission session (reverse link + forward link), UTs are categorized into two groups that defines which TDD protocol format should be used.

During reverse link

- UTs that have no previous known CSI at the BS or with low SNR regime during the previous downlink need to upload an initiative format, which is a conventional format to allow channel estimation at the BS.
- UTs that have the previous high SNR during last transmission session do not need to send their pilots, supposing that BS can predict their next CSI. This group of UTs uploads their data using predictive TDD protocol format.

During BS processing

- BS have to estimate CSI for UTs sending their pilots (uploading initiative format) and predict (using CSI Map) the CSI of other UTs (uploading predictive format).
- The uplink data can be estimated using a hybrid matrix of predicted QCSI and estimated CSI.

During Forward Link

- At the BS, precoding will take place using the hybrid channel matrix of estimated CSI and predicted QCSI.
- Each Downlink field is tagged with a test symbol (known at both BS and UTs) to allow UTs to easy compute their SNR (referring to a specific threshold) and decide whether to use predictive or initiative format at the next transmission session.

The BS can use the test symbol, as a control flag to oblige UTs to send their pilots. This can be done by reducing the associated power of the test symbol.

The estimated CSI from users sending initiative format will be quantized and used as a learning data to update the CSI map. Figure 2 shows the proposed system diagram that represents different operation between BS and UTs during the transmission session.

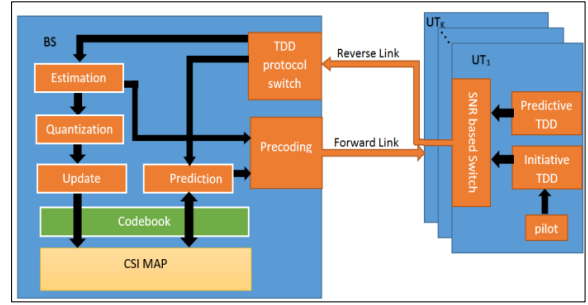

*Figure 2 Block diagram shows different UT and BS reciprocity operations*

Referring to figure 2, TDD protocol switch at the BS is used to switch between predictive TDD protocol formats and initiative formats based on the header content, where channel prediction and channel estimation will take place respectively. At the UT side, SNR based switch will decide which format to upload based on previously measured SNR of the (Test Symbol). At the BS, CSI responding to initiative TDD formats will be estimated and then will be sent in parallel to precoding and map update. A quantized version of the estimated CSI (QCSI) will be used to update the CSI map after translations using a codebook.

## IV. CHANNEL STATE INFORMATION MAP

We use graph theory and machine-learning approach to define a network of connected nodes. Each node $N_i$ stores a reference of a unique QCSI, estimated from previous transmission sessions. Nodes are connected with weighted directed connections called edges (i.e. $E_{i,x} = \omega$ refers to the edge directed from $N_i$ to $N_x$ with assigned weight $0 \leq \omega \leq 1$ that represents the frequency of this transition), see figure 3.

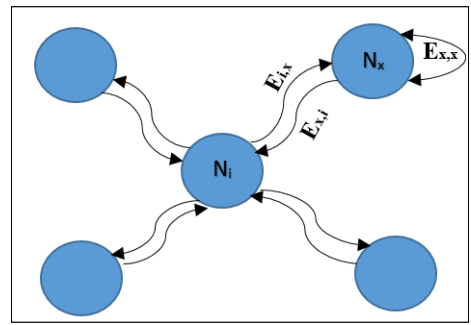

*Figure 3. CSI node map*

For example, if any UT uploads an initiative TDD format for two consecutive sessions, CSI map will add a node for each QCSI (if not exist) and update or create the edge corresponding to the transition from $N_i$ to $N_x$, where $N_i$ and $N_x$ correspond to the first QCSI and Next QCSI respectively. The weight of $E_{i,x}$ will increase as much as there is a transition from $N_i$ to $N_x$. In the case of no transition for several TDD sessions, an edge will be created from the current node to itself and will be updated as much as there is no transition.

The CSI map had to monitor the CSI of each UT in the cell and it is updated on each CSI estimation (see figure 4).

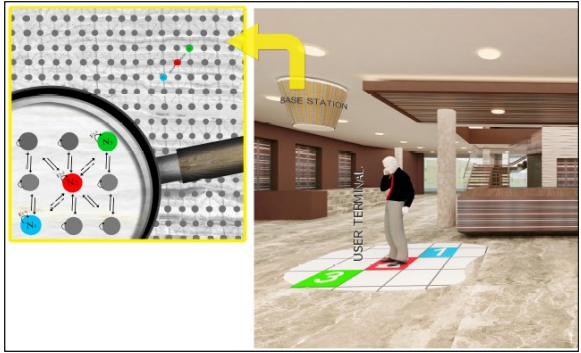

*Figure 4. Conception of CSI map representing an indoor scenario*

In order to control the huge amount of nodes that had been created, a garbage collection algorithm was developed. This algorithm works periodically to delete nodes that have weakly connected edges, in other words, edge weights below a specific threshold *TH* will be disconnected then free nodes will be deleted.

## V. CSI QUANTIZATION

The quantization of CSI in this paper is done at the BS and stored at the BS for later uses. QCSI parameters are stored in a two-part codebook $Z$ and $R$, where the first one stores a finite set of shadow fading parameters and the second one stores a finite set of distances between BS and UT.

We model the channel between the j[th] BS and the k[th] UT of the l[th] cell as $g_{jlk} = \sqrt{\beta_{jlk}}\, h_{jlmk}$,

i.e. $h_{jlmk} \triangleq [H_{jl}]_{m,k}$.

From (10), as M>> K we can ignore the effect of fast fading and simply write.

$$g_{jlk} = \sqrt{\beta_{jlk}}$$

The large-scale fading decomposes as $\beta_{jlk} = \frac{z_{jlk}}{r_{jlk}^\gamma}$, where $z_{jlk}$ represents the shadow fading with lognormal distribution with standard deviation $\sigma_{Shadow}$ and $\gamma$ is the path loss exponent.

Followed a method used in [13] and [5], we can divide the space of all possible channel realization of Z and R into $I, N$ vector length respectively.

Where $\hat{z}_i = \{Z : |z_i{}^2 - Z| < |z_j{}^2 - Z|\ for\ all\ j \neq i\ \}$ and $z_i$ scalar representing region $\hat{z}_i$, and $\hat{r}_n = \{R : |r_n{}^2 - R| < |r_j{}^2 - R|\ for\ all\ j \neq n\ \}$ where $r_n$ scalar representing region $\hat{r}_n$.

To design Z and R we use the classical non-uniform quantizer algorithm used in [5].

The parameters of Z and R can be acquired from estimated CSI by minimizing the root mean square error as follows:

$$\arg min_{z_i, r_n} \left[ \frac{1}{IN} \sum_{n=1, i=1}^{I,N} \sqrt{\hat{g}_{jlk}{}^2 - \frac{z_i}{r_n^\gamma}} \right]$$

Where $I, N$ are the length of the finite codebooks Z and R respectively, and $z_i$, $r_n$ are the i[th] and the n[th] elements of Z and R respectively, $\hat{g}_{jlk}$ is the estimated version of CSI related to user k.

## VI. CSI MAP LEARNING ALGORITHM

CSI learning algorithm corresponds to the systematic process of updating the directed edge weights and creating new nodes in the CSI map. The following flowchart of figure 5 presents the learning algorithm.

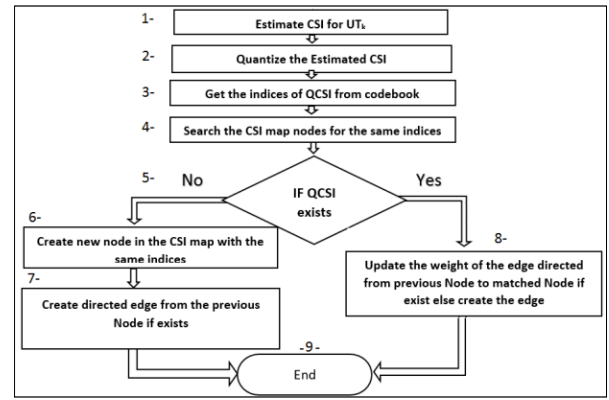

*Figure 5. Flowchart of CSI map learning algorithm*

Referring to the flowchart of figure 5, steps 1 and 2 will estimate then quantize CSI respectively. At step 3 a codebook will be used to find the indices of the QCSI, after that the algorithm will search for the existence of the estimated QCSI. In case the estimated QCSI is new, a node storing the indices of this QCSI will be created and an edge from the previous node to the current node will be created. If the estimated QCSI is already on the map, the weight of the edge issued from the previous node to the current node will be updated. Weight update will take place on all edges connected to the last node before the transition. We define two weight update equations as follows:

$$W_{c\,(new)} = W_c + \Theta$$
Subjected to $\Theta + W_c \leq 1$.

$$W_{c'\,(new)} = W_{c'} - \left(\frac{W_{c'}}{\sum_{c'=1,c'\neq c}^{F} W_{c'}}\right)\Theta \quad \text{Subjected to}$$
$W_{c'\,(new)} \geq 0$

Where $c$ corresponds to the index of the winning edge that represent the transition from the current node to the next node, $c'$ corresponds to the set of edge indices that issued from the current node and does not represent the transition. $F$ is the number of edges issued from the current node and $0 < \Theta < 1$ controls the learning speed.

The total weight of all edges issued from any node are limited as follows: $\sum_{c'=1}^{F} W_{c'} = 1$.

At prediction phase, the next possible CSI of $UT_k$ can be simply found by following the edge with maximum weight issued from the current node.

## VII. NUMERICAL RESULTS

In order to validate the performance of CSI map algorithm, we develop a simulation software, which is able to create and update CSI map at the BS. Our results consider a system with six cells each including one BS equipped with 420 antennas. The six cells cover an indoor area of 300 m$^2$, representing a floor in a hotel.

Each cell serves 30 UTs and shares an interference area of 15% of its cell area with neighbor cells. UTs are initially uniform normal distributed and moves randomly in the area. At each TDD session the new CSI estimated from UTs, are used to update the CSI map. The results presented in this paper ignores the error results from false CSI prediction.

Figure 6 shows an increase in the sum-rate with different prediction hit ratios compared to conventional Massive MIMO, which is represented by zero hit ratio.

The progression of the hit ratio will increase and goes toward stability as TDD sessions between BS and UTs increases. It is obvious from figure 7 that, the hit ratio is not stable before 50000 TDD sessions, which is due to the lack of CSI of new positions in the cell and it goes toward stability as much as the CSI map get more matured after several learning epochs.

The increase in the hit ratio is directly proportional to the amount of uploaded predictive TDD formats (1-α), which leads to less uploaded pilots or initiative TDD formats and thus, less uplink pilot contamination.

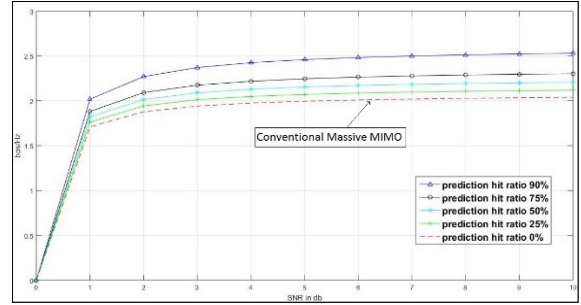

*Figure 6. Sum-rate vs. SNR with different prediction hit rate*

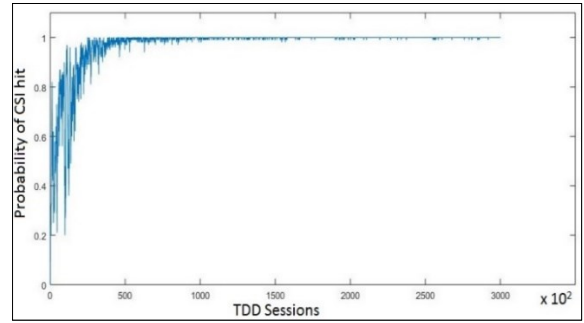

*Figure 7. Hit ratio with respect to the number of TDD sessions*

## VIII. CONCLUSION

We propose a technique to reduce uplink pilot feedback by predicting CSI at the BS instead of estimating CSI for each uplink session. Such a technique will have a major impact on mitigating uplink pilot contamination and increasing system sum-rate. Simulation results had proved the efficiency of mitigating pilot contamination using CSI Map approach. Where further researches on the implementation of CSI Map in an outdoor scenario will be studied.

# REFERENCES

[1] T. L. Marzetta, "How much training is required for multiuser MIMO?," *Conf. Rec. - Asilomar Conf. Signals, Syst. Comput.*, pp. 359–363, 2006.

[2] H. Q. Ngo, T. L. Marzetta, and E. G. Larsson, "Analysis of the pilot contamination effect in very large multicell multiuser MIMO systems for physical channel models," *ICASSP, IEEE Int. Conf. Acoust. Speech Signal Process. - Proc.*, vol. 2, no. 2, pp. 3464–3467, 2011.

[3] T. L. Marzetta, "Noncooperative cellular wireless with unlimited numbers of base station antennas," *IEEE Trans. Wirel. Commun.*, vol. 9, no. 11, pp. 3590–3600, 2010.

[4] J. Hoydis, S. ten Brink, and M. Debbah, "Massive MIMO: How many antennas do we need?," *2011 49th Annu. Allert. Conf. Commun. Control. Comput.*, pp. 545–550, 2011.

[5] B. Mielczarek and W. A. Krzymie, "Vector Quantization of Channel Information in Linear Multi-User MIMO Systems," pp. 302–306, 2006.

[6] B. Mielczarek and W. Krzymien, "Quantized channel state information prediction in multiple antenna systems," 2014.

[7] H. Shirani-Mehr, D. N. Liu, and G. Caire, "Channel state prediction, feedback and scheduling for a multiuser MIMO-OFDM downlink," *Conf. Rec. - Asilomar Conf. Signals, Syst. Comput.*, no. 1, pp. 136–140, 2008.

[8] D. Schafhuber and G. Matz, "MMSE and adaptive prediction of time-varying channels for OFDM systems," *IEEE Trans. Wirel. Commun.*, vol. 4, no. 2, pp. 593–602, 2005.

[9] J. Chang, I. T. Lu, and Y. Li, "Adaptive codebook based channel prediction and interpolation for multiuser MIMO-OFDM systems," *IEEE Int. Conf. Commun.*, no. October 2010, 2011.

[10] C. Li, J. Zhang, S. Song, and K. B. Letaief, "Selective Uplink Training for Massive MIMO Systems," no. 16211815, 2016.

[11] X. Wang, X. Wang, L. Gao, S. Mao, and S. Pandey, "DeepFi : Deep learning for indoor fingerprinting using channel state information DeepFi : Deep Learning for Indoor Fingerprinting Using Channel State Information," no. March 2016, 2015.

[12] F. Rusek, D. Persson, B. K. Lau, E. G. Larsson, T. L. Marzetta, O. Edfors, and F. Tufvesson, "Scaling up MIMO : Opportunities and challenges with very large arrays," *IEEE Signal Process. Mag.*, vol. 30, no. 1, pp. 40–60, 2013.

[13] Y. Linde, "An Algorithm for Vector Quantizer Design," vol. C, no. 1, pp. 84–95, 1980.

# ENDNOTES

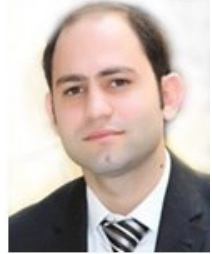

[1]**Ahmed A. Abboud** (2nd Oct 1985) was born in MazratMichref South Lebanon. Received a technical diploma in communication & computer engineering from University Institute of Technology (UIT) Jwaya, South Lebanon and the Master of Science in Computer Science & Communication from Arts, Sciences & Technology University in Lebanon, in 2008 and 2012 respectively. He is now a PHD student at the University of Limoges since October 2014.His research interests are in the area of applying artificial intelligence algorithms on MIMO communication systems.

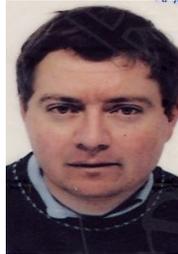

**JPC**[3] received the engineering degree from the Ecole Nationale Supérieure des Télécommunications de Bretagne in 1990. He also received the Aggregation degree in Physics in 1993 and the HDR degree in 2002 from the University of Limoges. He became a IEEE senior member in 2007. He is now a full professor in signal processing at the Ecole Nationale d'Ingénieurs de Limoges (ENSIL) since 2006. His research activities concern signal processing algorithms for digital communications including cooperative networks, network coding, space-time coding, and joint optimization of physical and MAC layers of future communication systems. He is getting involved in several French and European research programs.

**Dr. Vahid Meghdadi**[4] received the BSc and MSc degrees from Sharif University of Technology, Tehran, Iran, respectively in 1988 and 1991 and PhD degree from the University of Limoges, France in 1998. He has been working at the department of electronic and telecommunication of ENSIL/University of Limoges as assistant professor since 2000 and as full professor since 2014. He received in 2008 and 2012 from the French Ministry of Research and Higher Education the award of scientific excellence. His main interest in research is the telecommunication systems including MIMO systems, coding, network coding, cooperative communications, sensor network and smart grid. Since 1998, he has been scientific manager for more than 10 research projects in the field of ICT (information and Communications Technology). He is the (co-)author of more than 100 publications in scientific journals and conferences and served as TPC members in several international conferences.